\documentstyle[prl,aps,twocolumn]{revtex}
\begin{document}

{\large \bf Comment on ``Charged Particle Ratio Fluctuation as a Signal 
for Quark-Gluon \\ Plasma" and ``Fluctuation Probes \\ 
of Quark Deconfinement"}

\vspace{0.5cm}

Charge fluctuations studied on event-by-event basis have been suggested 
by two groups \cite{Jeo00,Asa00} (see also \cite{Fia00})
to provide a signal of the quark-gluon plasma (QGP) produced in heavy-ion 
collisions at high energies. Specifically, Jeon and Koch \cite{Jeo00} and 
Asakawa, Heinz and M\"uller \cite{Asa00} proposed to study the fluctuation 
measure 
\begin{equation}\label{measure1}
{V(Q) \over \langle N_{\rm ch} \rangle} = 
{\langle  (Q - \langle Q \rangle )^2 \rangle 
\over \langle N_{\rm ch} \rangle }\;,
\end{equation}
where $Q$ is the charge (electric or baryon) and $N_{\rm ch}$ is the number 
of electrically charged particles both measured in a given rapidity interval; 
$\langle ... \rangle$ denotes averaging over events. The measure 
(\ref{measure1}) equals approximately unity for the equilibrium hadron gas 
and it is 2 to 3 times smaller \cite{Jeo00,Asa00} when the charge fluctuations 
are generated in QGP and the entropy, which controls the final state particle 
multiplicity, is conserved. Since various phenomena which occur in the hadron 
phase rather increase than decrease the fluctuations created in QGP the 
smallness of (\ref{measure1}), if measured, seems to be an unambiguous signal 
of the plasma production. 

Jeon and Koch \cite{Jeo00} and Asakawa {\it et al.} \cite{Asa00} assume 
that the final state charge fluctuations are created when QGP produced at 
the collision early stage achieves local thermodynamic equilibrium. They 
argue that the fluctuations are not increased to the level characteristic 
for the hadron phase due to the rapid system expansion. However, it seems 
a legitimate assumption that the fluctuations are determined even earlier 
i.e. at the initial stage of heavy-ion collision when the energy of the 
participating nucleons is released and nonequilibrium parton system emerges. 
The longitudinal collective motion, which counteracts the relaxation of 
the fluctuations, develops just at this stage. Thus, we assume that the 
final state charge fluctuations observed in a rapidity window $\Delta y_f$ 
are fully determined by the initial stage charge fluctuations in the 
rapidity window $\Delta y_i$, which corresponds to $\Delta y_f$. Such 
a scenario does not exclude formation of the equilibrium QGP. It only 
demands that the processes of thermalization, hadronization, etc. 
redistribute the charges at the scale smaller than $\Delta y$. Then, 
the charge within the window is conserved during the temporal evolution.

The number of charge carriers dramatically increases at the collision 
early stage. If the charge fluctuations are indeed created initially 
the numerator of the measure (\ref{measure1}) is determined by the much 
smaller number of carriers than the denominator. Therefore, the measure 
is expected to be {\em small}. One gets a rough estimate of (\ref{measure1}) 
assuming that the initial fluctuations of the electric charge in $\Delta y_i$ 
are due to the variation of the number of protons occurring in $\Delta y_i$.  
For the central collisions and $\Delta y_i$ being nuch smaller than the full
rapidity interval, these fluctuations are  expected to be poissonian.  
Therefore, the measure (\ref{measure1}) reads 
\begin{equation}\label{measure2}
{V(Q) \over \langle N_{\rm ch} \rangle } = 
{\langle Z \rangle \over \langle N_{\rm ch} \rangle } \;,
\end{equation}
where $Z$ is number of participating protons. In the case of baryon charge 
fluctuations one replaces $Z$ by the number of participating nucleons. 
In order to remove a possible contribution of the geometrical fluctuations 
it was proposed in \cite{Jeo00} to use 
\begin{equation}\label{measure3}
{1 \over 4} \langle N_{\rm ch} \rangle V(N^+/N^-) \;, 
\end{equation}
instead of (\ref{measure1}) in experimental studies of the electric 
charge fluctuations. The measure (\ref{measure3}) approximately equals 
(\ref{measure1}) for sufficiently small multiplicity fluctuations. If the
charge fluctuations are indeed governed by the initial ones the use of 
(\ref{measure3}) instead of (\ref{measure1}) will reduce the measured 
fluctuations even in the central collisions because the charge fluctuations 
which are proportional to the produced entropy do not contribute 
to (\ref{measure3}).    

Since the charge particle multiplicity per participating proton is 
about 8 at CERN SPS energy \cite{NA49} and it is even larger at BNL RHIC, 
the value given by eq.~(\ref{measure2}) is significantly smaller than 
unity. One notes that our estimate remains similar if the valence quarks 
instead of protons are the charge carriers at the collision early stage. 
Thus, the smallness of charge fluctuations, if measured, can be 
interpreted either as a result of the equilibrium  QGP or due to the  
initial charge variation. Rapid expansion of the matter produced in  
nucleus-nucleus collisions can freeze both types of fluctuations or the 
superposition of them. If the mechanism of freezing is not efficient 
enough the fluctuations characteristic for the hadronic final state
will be observed.

\vspace{0.2cm}

Marek Ga\'zdzicki \\
Institut f\"ur Kernphysik, Universit\"at Frankfurt \\
August--Euler--Strasse 6, D - 60486 Frankfurt, Germany

\vspace{0.2cm}

Stanis\l aw Mr\' owczy\' nski \\
So\l tan Institute for Nuclear Studies \\
ul. Ho\.za 69, PL-00-681 Warsaw, Poland \\



\end{document}